\documentclass[aps,prl,reprint,superscriptaddress,amsmath,amssymb]{revtex4-2}

\usepackage{dcolumn}
\usepackage{bm}
\usepackage{graphicx}
\usepackage{amssymb}
\usepackage{amsmath, amscd}  
\newcommand{\R}{\mathbb{R}}

\newenvironment{itquote}
  {\begin{quote}\itshape}
  {\end{quote}\ignorespacesafterend}

\begin{document}

\title{Quantum Paradoxes and the Quantum-Classical Transition under Unitary Measurement Dynamics with Random Hamiltonians}

\author{Alexey A. Kryukov}
\affiliation{Department of Mathematics \& Natural Sciences, University of Wisconsin-Milwaukee, USA}
\email{kryukov@uwm.edu}

\begin{abstract}
We develop a dynamical framework for quantum measurement based on stochastic but unitary evolution in projective state space. Random Hamiltonians drawn from the Gaussian Unitary Ensemble generate stochastic unitary dynamics of the quantum state, while equivalence classes reflecting finite detector resolution define classical observables as well as classical configuration-space and phase-space submanifolds. When the evolution is constrained to the phase-space submanifold, free Schr\"odinger dynamics reduces to Newtonian motion, while stochastic motion constrained to the classical configuration-space submanifold yields ordinary Brownian motion in classical space. Transition probabilities under the stochastic dynamics satisfy the Born rule, whereas the constrained classical evolution produces the normal probability distributions characteristic of classical measurements. We show that, in this setting, measurement, state reduction, and the quantum-classical transition emerge from unitary dynamics alone, without invoking nonunitary collapse or additional postulates. Entanglement and EPR correlations arise geometrically from the evolution of joint states in composite state space, preserving locality in spacetime. The framework provides a unified dynamical account of measurement and classicality compatible with the structure of quantum mechanics.
\end{abstract}

\maketitle

\section{Introduction}

Despite the extraordinary empirical success of quantum mechanics, its conceptual foundations remain unsettled.
Central difficulties cluster around measurement: the emergence of definite outcomes from linear, unitary dynamics; the role of observers and environments; and the appearance of classical trajectories. Many well-known paradoxes and foundational problems, including wave-function collapse, Schr\"odinger's cat, Wigner's friend, nonlocal correlations, and the quantum-classical transition, are likewise rooted in the transition of quantum states under measurement.

Existing approaches to these problems typically take one of several routes: they either deny the physical reality of collapse and represent measurement by a mathematical projection; eliminate collapse in favor of infinite branching; assume permanently definite positions guided nonlocally by the wave function; or introduce nonlinear stochastic modifications of the Schr\"odinger equation to account for spontaneous collapse in position, energy, or other observables. The latter approach must reconcile linear unitary evolution with nonlinear state reduction while remaining consistent with observations, which severely restricts the admissible range of model parameters \cite{Donadi2021,Piscicchia2024,Bilardello2016,Vinante2015,Carlesso2016,Altamura2025,Adler2007,AdlerBassi2009,IGMHeating2017}.

Here, we propose an alternative approach that treats measurement as a physical interaction between a quantum system and its environment. The central conjecture is that the highly complex, rapidly fluctuating interactions with a measuring apparatus or environment can be modeled by a time-dependent random Hamiltonian drawn from an appropriate ensemble. The resulting dynamics is a linear stochastic evolution in state space governed by the Schr\"odinger equation. When classical space and phase space are identified as submanifolds of state space and states indistinguishable by detectors are grouped into equivalence classes, this framework avoids the standard objections to linear collapse models and provides a unified account of state reduction, classical behavior, and measurement outcomes \cite{KryukovNew,KryukovMath,KryukovPhysLett,Kryukov2020,KryukovPhysicsA,KryukovDICE24}.

The purpose of the present paper is not to rederive the formalism in detail, but to show how the resulting single dynamical framework sheds new light on a broad range of quantum paradoxes. We argue that many apparent inconsistencies of quantum theory arise from overlooking the role of the geometry of state space under stochastic unitary evolution, and from conflating exact quantum states with experimentally accessible information.
Under Schr\"odinger evolution with a random Hamiltonian, phenomena traditionally labeled as {\em collapse}, {\em branching}, or {\em observer dependence} arise as dynamical features of state-space trajectories conditioned on measurement records, whenever the evolving state intersects the classical submanifold of state space.

We systematically revisit several foundational paradoxes, including the measurement problem, wave function collapse, Born rule, outcome uniqueness, Quantum-to-Classical Transition, Schr\"odinger's cat, Wigner's friend, and aspects of nonlocality, from this perspective. In each case, the paradox is substantially clarified without invoking nonunitary dynamics, hidden variables, or observer-dependent postulates. Instead, the same mechanism, unitary evolution in state space driven by random Hamiltonians together with the use of equivalence classes of states, applies uniformly across microscopic and macroscopic regimes.

The paper is organized as follows. We begin with a brief review of the geometry of the projective state space, its classical configuration-space and phase-space submanifolds, the role of equivalence classes of states, and the evolution of states under random Hamiltonians. We then explain how this framework leads to a dynamical description of measurement and the emergence of the Born rule. Finally, we analyze a range of quantum paradoxes, showing how they are resolved within this framework, and summarize the overarching implications of the approach.

\section{The Formalism}

Realistic position-measuring devices cannot distinguish states that are sufficiently well localized in space. Consequently, whenever a state $\varphi$ has spatial support narrower than the instrumental resolution $\sigma$, or is negligible outside a region of spatial extent $\sigma$, the measurement identifies it as a position eigenstate.

Let $\varphi \in L_2(\mathbb{R}^3)$ be an arbitrary quantum state, written in the polar form
\[
\varphi(\mathbf{x}) = r(\mathbf{x})\,e^{i\Theta(\mathbf{x})}.
\]
Assume that $r$ has finite variance, and define its translated and rescaled versions by
\begin{equation}
\label{sigmao}
r_{\mathbf{a},\sigma}(\mathbf{x})
= \sigma^{-3/2}\,
r\!\left(\frac{\mathbf{x}-\mathbf{a}}{\sigma}\right).
\end{equation}
The parameter $\sigma$ represents the spatial resolution of the detector.
For small $\sigma$, any sufficiently smooth state localized near $\mathbf{a}$, viewed projectively in $\mathbb{CP}^{L_2}$, takes the approximate form
\begin{equation}
\label{phii}
\varphi(\mathbf{x})
= r_{\mathbf{a},\sigma}(\mathbf{x})\,e^{i\mathbf{p}\cdot\mathbf{x}/\hbar},
\end{equation}
since quadratic and higher-order terms in the Taylor expansion of $\Theta$ are negligible over the region where $r_{\mathbf{a},\sigma}$ is not negligibly small.

Let $M^{\sigma}_{3,3}\subset\mathbb{CP}^{L_2}$ denote the set of wave packets of the form \eqref{phii} for a fixed function $r$. The parameters ${\bf a}$ and ${\bf p}$ represent the approximate position and momentum of the packet, with the momentum defined via group velocity.
Equipped with the metric induced from the Fubini-Study metric on $\mathbb{CP}^{L_2}$, this set becomes a Riemannian manifold.
With an appropriate choice of units, the map
\[
\Omega:\;(\mathbf{a},\mathbf{p})
\longmapsto
r_{\mathbf{a},\sigma}\,e^{i\mathbf{p}\cdot\mathbf{x}/\hbar}
\]
is an isometry between the Euclidean phase space $\mathbb{R}^3\times\mathbb{R}^3$ and $M^{\sigma}_{3,3}$.
Via $\Omega$, $M^{\sigma}_{3,3}$ may also be endowed with a linear structure inherited from $\mathbb{R}^3\times\mathbb{R}^3$.

Consider now the action functional
\begin{equation}
\label{SS}
S[\varphi]
=
\int
\overline{\varphi}(\mathbf{x},t)
\left[
i\hbar\frac{\partial}{\partial t}
-
\widehat h
\right]
\varphi(\mathbf{x},t)\,
d^3\mathbf{x}\,dt,
\end{equation}
where
\begin{equation}
\label{h}
\widehat h
=
-\frac{\hbar^2}{2m}\Delta
+
\widehat V(\mathbf{x},t).
\end{equation}
Unconstrained variation of $S[\varphi]$ yields the Schr\"odinger equation.
If, however, $\varphi$ is constrained to lie on $M^{\sigma}_{3,3}$ with sufficiently small $\sigma$, the action reduces to
\begin{equation}
\label{SSS}
S
=
\int
\left[
\mathbf{p}\frac{d\mathbf{a}}{dt}
-
h(\mathbf{p},\mathbf{a},t)
\right]dt,
\end{equation}
with
\begin{equation}
\label{h1}
h(\mathbf{p},\mathbf{a},t)
=
\frac{\mathbf{p}^2}{2m}
+
V(\mathbf{a},t).
\end{equation}
Constrained variation yields Newton's equations, in agreement with the Ehrenfest theorem for narrow wave packets.
Thus, within this framework, a Newtonian particle corresponds to a quantum system whose state is constrained to $M^{\sigma}_{3,3}$. The manifold $M^{\sigma}_{3,3}$ may therefore be identified with the classical phase space of the particle.

This construction extends directly to many-body systems.
For example, a two-particle state constrained to
$M^{\sigma}_{3,3}\otimes M^{\sigma}_{3,3}$ and evolving under
\begin{equation}
\label{twoh}
\widehat h
=
-\frac{\hbar^2}{2m_1}\Delta_1
-
\frac{\hbar^2}{2m_2}\Delta_2
+
\widehat V(\mathbf{x}_1,\mathbf{x}_2,t)
\end{equation}
follows classical Newtonian dynamics.

The freedom in choosing $r$ allows $M^{\sigma}_{3,3}$ to be defined via equivalence classes of sufficiently localized states.
For concreteness, we adopt Gaussian representatives,
\begin{equation}
\label{g}
g_{\mathbf{a},\sigma}(\mathbf{x})
=
\left(\frac{1}{2\pi\sigma^2}\right)^{3/4}
\exp\!\left[-\frac{(\mathbf{x}-\mathbf{a})^2}{4\sigma^2}\right].
\end{equation}
For states $\varphi=g_{\mathbf{a},\sigma}e^{i\mathbf{p}\cdot\mathbf{x}/\hbar}$,
the Schr\"odinger velocity
\[
\frac{d \varphi}{d t}
=
-\frac{i}{\hbar}\widehat h\varphi
\]
decomposes orthogonally in the Fubini-Study metric into components corresponding to classical velocity, classical acceleration, and wave-packet spreading.
Their squared norms sum to
\begin{equation}
\label{decomposition}
\left\|\frac{d\varphi}{dt}\right\|^2_{\mathrm{FS}}
=
\frac{\mathbf{v}^2}{4\sigma^2}
+
\frac{m^2\mathbf{w}^2\sigma^2}{\hbar^2}
+
\frac{\hbar^2}{32m^2\sigma^4},
\end{equation}
where $\mathbf{v}=\frac{d\mathbf{a}}{dt}$ and $\mathbf{w}=-\nabla V/m$.
Suppressing the spreading term constrains the motion to $M^{\sigma}_{3,3}$, reducing commutators to Poisson brackets and yielding classical dynamics \cite{KryukovNew}.

The isometry $\Omega$ relates Euclidean distances in classical phase space
to Fubini-Study distances in state space.
For
$\varphi=g_{\mathbf{a},\sigma}e^{i\mathbf{p}\cdot\mathbf{x}/\hbar}$ and
$\psi=g_{\mathbf{b},\sigma}e^{i\mathbf{q}\cdot\mathbf{x}/\hbar}$,
one finds
\begin{equation}
\label{mainOO}
e^{-\frac{(\mathbf{a}-\mathbf{b})^2}{4\sigma^2}
-\frac{(\mathbf{p}-\mathbf{q})^2}{\hbar^2/\sigma^2}}
=
\cos^2\rho(\varphi,\psi),
\end{equation}
where $(\mathbf{a}-\mathbf{b})^2$ and $(\mathbf{p}-\mathbf{q})^2$ denote squared Euclidean distances in $\R^3$.
Restricting $\Omega$ to position space yields an isometry
$\omega:\mathbf{a}\mapsto g_{\mathbf{a},\sigma}$
between $\mathbb{R}^3$ and the submanifold
$M^{\sigma}_3\subset\mathbb{CP}^{L_2}$,
characterized by
\begin{equation}
\label{mainO}
e^{-\frac{(\mathbf{a}-\mathbf{b})^2}{4\sigma^2}}
=
\cos^2\rho(g_{\mathbf{a},\sigma},g_{\mathbf{b},\sigma}).
\end{equation}
This identification can be used to demonstrate that the normal distribution on $\mathbb{R}^3=M^{\sigma}_3$
corresponds exactly to, and uniquely extends as, the Born rule in $\mathbb{CP}^{L_2}$ \cite{KryukovPhysicsA}.

Classically, a position measurement can be modeled as a random walk on $\mathbb{R}^3$, approximating Brownian motion over the time interval of observation. Such a model is physically well motivated, since measurement errors arise from the cumulative effect of many small fluctuations produced by interactions between the particle, the measuring device, and the surrounding environment.

We assume that the situation for position measurements of microscopic particles is analogous. Rapidly fluctuating interactions between the particle, the measuring device, and the environment make applicable the reasoning originally introduced by Wigner \cite{Wigner} and later formalized in the Bohigas-Giannoni-Schmit conjecture \cite{BGS}, now in a time-dependent measurement setting.
Imposing, in addition, Hermiticity of the Hamiltonian governing the evolution, we propose the following conjecture:
\begin{itquote}{\bf (RM)}
During position measurement, the state evolves via a random walk in state space.
In the absence of drift, each step is generated by Schr\"odinger evolution with a Hamiltonian independently drawn from the Gaussian Unitary Ensemble.
\end{itquote}
It can be shown that this walk provides a unique extension of the Gaussian random walk on $\mathbb{R}^3=M^{\sigma}_3$ to a random walk on $\mathbb{CP}^{L_2}$ whose step distribution is invariant under unitary transformations \cite{KryukovNew,Kryukov2020,KryukovPhysicsA}. In this sense, the assumptions underlying conjecture {\bf (RM)} are similar to those employed by Einstein in his theory of Brownian motion \cite{Ein}.

Under {\bf (RM)}, the transition probability depends only on the Fubini-Study distance between states. When the walk is restricted to $M^{\sigma}_3$, this probability agrees with a Gaussian distribution. Consequently, the induced transition probability in the full state space is given by the Born rule \cite{KryukovPhysicsA}.
The apparent concern that a state might never reach $M^{\sigma}_3$ is resolved once $M^{\sigma}_3$ is defined in terms of equivalence classes of detector-indistinguishable states, as we now demonstrate.

In one dimension, let
\begin{equation}
\label{g1}
g_{a,\sigma}(z)
=
\left(\frac{1}{2\pi\sigma^2}\right)^{1/4}
\exp\!\left[-\frac{(z-a)^2}{4\sigma^2}\right].
\end{equation}
The set of all such functions forms a submanifold $M^{\sigma}_1$ of the state
space $\mathbb{CP}^{L_2}$ with the induced metric, which is Euclidean, making $M^{\sigma}_1$ isometric to $\R$.

The equivalence class $\{g_c\}$ consists of all states with position expectation
$\mu_z=c$ and standard deviation $\delta_z\le\sigma$.
Distances between a state and an equivalence class are defined by
\begin{equation}
\label{dist}
\rho(\varphi,\{g_c\})
=
\inf_{\psi\in\{g_c\}}\rho(\varphi,\psi),
\end{equation}
and the distance between two equivalence classes by
\begin{equation}
\label{dist1}
\rho(\{g_c\},\{g_d\})
=
\inf_{\varphi\in\{g_c\}}\rho(\varphi,\{g_d\}).
\end{equation}

In the latter case, the infimum is realized by the Gaussian representatives
$g_{c, \sigma}$ and $g_{d, \sigma}$ in (\ref{g1}), which minimize the Fubini-Study distance among all states in
the corresponding equivalence classes.
Thus the space $\widetilde M^{\sigma}_1$ of equivalence classes is itself a Riemannian manifold, isometric to
$M^{\sigma}_1$ and hence to $\mathbb{R}$.
Each equivalence class is large, absorbing infinitely many orthogonal functions in the Hilbert space.
Augmenting equivalence classes with momentum factors yields the manifold $\widetilde M^{\sigma}_{1,1}$, equipped with the induced Euclidean metric of $\mathbb{R}^2$. As before, Schr\"odinger evolution constrained to $\widetilde M^{\sigma}_{1,1}$ reduces to Newtonian dynamics.

In this setting, the position of a particle is specified by an equivalence class of states characterized by the parameters $\mu_z$ and $\delta_z$, with all remaining degrees of freedom absorbed into the class. Translating and scaling any suitable initial state $\varphi$ generates a two-dimensional submanifold $M_\varphi\subset\mathbb{CP}^{L_2}$ with orthogonal coordinates $(\tau,s)=(\mu_z,\ln\delta_z)$. State reduction is then described by a random walk on $\mathbb{R}^2=M_\varphi$.
The $\tau$-component of the walk yields a normal probability distribution, which extends uniquely to the Born rule on state space, while the 
$s$-component yields a probability of order $1/2$ for satisfying $\delta_z\le\sigma$, accounting both for the emergence of collapse and for the persistence of classical behavior \cite{KryukovPhysicsA}.

\section{Quantum Paradoxes and Quantum-Classical Transition}

\subsection{Measurement Problem}

\emph{Question:} How can linear Schr\"odinger evolution yield definite outcomes governed by the Born rule?

\medskip

\emph{Answer:}
For a single particle, the Schr\"odinger evolution driven by the random Hamiltonian in {\bf (RM)} induces a random walk on the projective state space $\mathbb{CP}^{L_2}$.  
Whether the initial state is a superposition of position states is immaterial: transition probabilities depend only on the Fubini-Study distance between states.  
After sufficiently many steps, the probability that the evolving state lies on the classical space submanifold $\widetilde M^{\sigma}_1$, i.e., satisfies $\delta_z \le \sigma$, approaches $1/2$.  
Conditioned on the state being on $\widetilde M^{\sigma}_1$, the probability of reaching a particular equivalence class $\{g_c\}$, corresponding to registering the position $z=c$, is given by the Born rule \cite{KryukovPhysicsA}.

\medskip

For a two-particle system in an arbitrary initial state, the Hilbert space is the tensor product
\[
L_2(\mathbb{R})\otimes L_2(\mathbb{R}) \cong L_2(\mathbb{R}^2),
\]
with the corresponding projective state space $\mathbb{CP}^{L_2}$.  
The classical space submanifold is the tensor product
\[
\widetilde M^{\sigma_1}_1 \otimes \widetilde M^{\sigma_2}_1,
\]
with possibly different resolution parameters $\sigma_1$ and $\sigma_2$ for the two particles.
The metric relation (\ref{mainO}) generalizes to
\begin{equation}
\label{mainOOO}
e^{-\frac{(a-b)^2}{4\sigma_1^2}-\frac{(c-d)^2}{4\sigma_2^2}}
=
\cos^2\rho (
g_{a,\sigma_1}\!\otimes\! g_{c,\sigma_2},
g_{b,\sigma_1}\!\otimes\! g_{d,\sigma_2}).
\end{equation}

To obtain an isometry between the Euclidean space $\mathbb{R}^2$ of position pairs $(a,c)$ and the manifold
$\widetilde M^{\sigma_1}_1 \otimes \widetilde M^{\sigma_2}_1$, equipped with the induced Fubini-Study metric, distances must be measured in units of $\sigma_1$ and $\sigma_2$ for the first and second particles, respectively; in general, the coordinate units along the two axes of $\R^2$ therefore differ.  
With this identification, the isotropic random walk of the joint state described by {\bf (RM)} induces a random walk on $\mathbb{R}^2$ corresponding to two independent Brownian motions along the coordinate axes, with diffusion coefficients set by $\sigma_1$ and $\sigma_2$.
In particular, by an appropriate choice of these parameters, one can decrease or increase the effect of measurement on the particles. This flexibility will be important later, when considering a system consisting of a particle coupled to a macroscopic measuring device.

The isotropy and homogeneity of the step distribution in {\bf (RM)} ensure that transition probabilities depend only on the Fubini-Study distance between joint states, and the Born rule follows exactly as in the single-particle case.  
Correlations between measurement outcomes for entangled particles follow directly from the Born rule within this framework. No additional nonlocal mechanisms beyond those encoded in the Hamiltonian dynamics of {\bf (RM)} are required.  
The extension to systems of more than two particles proceeds analogously.

\subsection{Wave Function Collapse}

\emph{Question:} What is wave function collapse, and why is the process stochastic despite deterministic Schr\"odinger dynamics? How can it appear effectively instantaneous and yield a single outcome, despite an initial superposition of possible outcomes?

\medskip

\emph{Answer:}
Stochastic outcomes arise from the random Hamiltonian drawn from the Gaussian Unitary Ensemble in {\bf (RM)}. What is traditionally called ``collapse'' corresponds here to the dynamical approach of the state toward the classical space submanifold $\widetilde M^{\sigma}_1$, or, for composite systems, toward a tensor product of such manifolds. For clarity, we restrict the discussion to a single particle, but the same mechanism applies to multiparticle states, as already noted in the discussion of the measurement problem.

Once the evolving state reaches a particular equivalence class $\{g_c\}\in\widetilde M^{\sigma}_1$, the particle's position is uniquely defined and can be registered with certainty. By choosing the time step of the stochastic evolution sufficiently small, the duration of this approach can be made arbitrarily short, accounting for the effectively instantaneous character of collapse observed in practice.

Outcome uniqueness follows directly from this dynamics. At any given time the state approaches a single equivalence class $\{g_c\}$ rather than branching into multiple simultaneously realized alternatives. The stochasticity of the evolution determines which equivalence class is reached, but once reached, the outcome is definite and unambiguous. In this framework, outcome uniqueness is therefore a dynamical consequence of the evolution rather than an additional postulate.

It is important to emphasize that the act of recording the fact that the state lies in $\{g_c\}$ is not part of the collapse process itself. Recording neither influences the evolution of the state nor, by itself, contributes to the stability of the outcome. If a second position measurement is performed immediately after the first, the same result will be obtained simply because the state has not yet had time to move appreciably away from $\{g_c\}$ under the dynamics.
To understand how recording becomes possible in physical terms, we now turn to the quantum-classical transition.

\subsection{Quantum-Classical Transition}

\emph{Question:}
How can Newtonian mechanics emerge if Schr\"odinger dynamics is universal? Why do macroscopic superpositions not persist?

\medskip

\emph{Answer:}
Macroscopic bodies continuously interact with particles and radiation in their environment. Scattered environmental particles and radiation carry information about a body's position, making the conjecture {\bf (RM)} applicable in this setting. Since Schr\"odinger evolution of a particle whose state is constrained to $\widetilde M^{\sigma}_{1,1}$ is Newtonian, the central issue is to explain why the spreading component in (\ref{decomposition}) is effectively suppressed during environmental monitoring.

Empirically, wave-function collapse appears nearly instantaneous, so that the contribution of the free Hamiltonian during individual random steps is negligible. We therefore assume that, during measurement kicks, the random Hamiltonian $\widehat h_{\mathrm{RM}}$ dominates the free Hamiltonian $\widehat h$. In this regime, the evolution of state of a macroscopic body under environmental measurement consists of short intervals of free Schr\"odinger evolution interspersed with rapid, stochastic ``kicks'' generated by $\widehat h_{\mathrm{RM}}$, during which the effect of $\widehat h$ can be neglected. The total Hamiltonian remains
$\widehat h_{\mathrm{tot}}=\widehat h+\widehat h_{\mathrm{RM}}$, while the dynamics effectively factorizes into alternating segments of free evolution and stochastic unitary steps.

Suppose that the initial state of a macroscopic particle lies on $\widetilde M^{\sigma}_1$. As in ordinary Brownian motion, the properties of the particle and its environment determine the time-step $dt$ and step-size $dz$ parameters of the random walk in {\bf (RM)}. For a macroscopic body in a natural environment (air, radiation), we assume that the time step is extremely small, while diffusion is strongly suppressed, so that the diffusion coefficient
$D=(dz)^2/dt$ remains small. As a result, many steps of the walk occur within a short time interval $\Delta t$, and neither the free Schr\"odinger evolution nor the stochastic component of the dynamics drives the state far from a small neighborhood of $\widetilde M^{\sigma}_1$.

Over $\Delta t$, in agreement with  \eqref{decomposition}, the state undergoes a Newtonian displacement along $\widetilde M^{\sigma}_1$, together with only a small excursion away from the manifold due to Schr\"odinger spreading and the random walk generated by {\bf (RM)}. The large number of steps further ensures that the state repeatedly returns to $\widetilde M^{\sigma}_1$, that is, satisfies the condition $\delta_z\le\sigma$, with probability arbitrarily close to one. At any given time, the probability of satisfying this condition is approximately $1/2$. Whenever this occurs, the position of the state on $\widetilde M^{\sigma}_1$ is recorded, and the stochastic evolution effectively restarts from that point, producing a sequence of recorded points on $\widetilde M^{\sigma}_1$.

This behavior is particularly transparent when described in terms of the orthogonal coordinates $(\mu_z,\ln\delta_z)$ on the plane $\mathbb{R}^2=M_\varphi$ \cite{KryukovPhysicsA}. The outcome of this process is a sequence of recorded positions of the state on $\R=\widetilde M^{\sigma}_1$, normally distributed about the Newtonian trajectory. This mechanism explains how Newtonian dynamics of a particle emerges from Schr\"odinger evolution in state space and, in particular, accounts for the classical behavior of macroscopic measuring devices.

In a system consisting of a measured particle and a macroscopic measuring device, the {\bf (RM)}-generated motion of the joint particle-device state constrained to $\widetilde M^{\sigma_p}_{1} \otimes \widetilde M^{\sigma_d}_{1}$ is described by two independent Brownian motions. The diffusion coefficients for these motions are determined by the spread parameters $\sigma_p$ and $\sigma_d$ of the particle and the device, respectively. The macroscopicity of the device implies that $\sigma_d \ll \sigma_p$ is extremely small, so that the state of the device, defined when the total state lies on $\widetilde M^{\sigma_p}_{1} \otimes \widetilde M^{\sigma_d}_{1}$, is effectively fixed on $\widetilde M^{\sigma_d}_{1}$.

The macroscopicity of the device allows us to assume that its state initially
lies in an equivalence class of $\widetilde M^{\sigma_d}_1$. Using the Gaussian
representative $\Psi \in M^{\sigma_d}_1$ of this class, the initial state of the
particle-device system is $\varphi \otimes \Psi$, where
$\varphi \in L_2(\mathbb{R})$ represents the state of the particle. Assume that
the parameters $dt$ and $dz$ of the random walk in {\bf (RM)} are sufficiently
small. The value of the parameter $\sigma_d$ will be specified
shortly.

Since the set of all translations of $\Psi$ is complete in $L_2(\mathbb{R})$, the state of the system after a single step of the stochastic evolution can be
written in the form
\begin{equation}
\label{Phi1}
\Phi_1 = \sum_k c_k\, f_k \otimes \Psi_k,
\end{equation}
where the unit-normalized functions $f_k$ are associated with the measured particle and the states $\Psi_k$ are translations of $\Psi$. By
choosing $\sigma_d$ sufficiently small, the state $\Phi_1$ can be approximated
arbitrarily well by a sum of the form \eqref{Phi1} in which the states
$\Psi_k$ are nearly orthogonal to one another, with exponentially small
pairwise inner products. If there exists at least one index $k$ such that
$\Psi_k$ is a nontrivial translation of $\Psi$ and the corresponding
coefficient $c_k$ is not negligibly small, then the Fubini-Study distance
between the initial state $\Phi=\varphi\otimes\Psi$ and the state $\Phi_1$ cannot be small, since $\Psi_k$ is nearly orthogonal to $\Psi$.

Consequently, for sufficiently small step-size $dz$, such superpositions are dynamically suppressed, and the state
after the first step must have the product form
$
\Phi_1 = \varphi_1 \otimes \Psi
$
for some particle state $\varphi_1$.
Since the same reasoning applies at every step of the evolution, and since the
number of steps required to complete the measurement is almost surely finite,
it follows that for sufficiently small $\sigma_d$ the state of the system retains the product form $\varphi_k \otimes \Psi$ at all times
during the measurement process with probability arbitrarily close to $1$.

Owing to the product form of the particle-device state, the state of the particle remains well defined throughout the measurement process and can be treated independently.
When, under {\bf (RM)}, the particle's state reaches the submanifold $\widetilde M^{\sigma_p}_{1,1}$, the joint particle-device state belongs to $\widetilde M^{\sigma_p}_{1,1} \otimes \widetilde M^{\sigma_d}_{1,1}$. On this manifold, the Hamiltonian (\ref{twoh}) reduces to the classical Hamiltonian of a pair of particles, and the measurement result is recorded as an ordinary Newtonian correlation between two classical subsystems, without invoking a collapse postulate or observer-dependent rules.

\subsection{Schr\"odinger's cat}

\emph{Question:}
How can quantum theory allow a macroscopic system, such as a cat, to exist in a superposition of classically distinct states like ``alive" and ``dead"?

\medskip

\emph{Answer:}
A macroscopic system such as a cat continuously interacts with its environment.
As in the particle-device system discussed in the previous section, the state of the atom-cat system at any time has the product form $\varphi \otimes \Psi$, where $\varphi$ is the atom's state and $\Psi \in \widetilde M^{\sigma}_{3,3}$ is the state of the cat.  
The initial atom-cat state evolves into either a classical ``alive" trajectory, culminating in the state  $\varphi_{{\rm undecayed}}\otimes \Psi_{{\rm alive}}$, or a classical ``dead" trajectory, ending in the state $\varphi_{{\rm decayed}}\otimes \Psi_{{\rm dead}}$, depending on the decay outcome, but never into a superposition of the two. The putative superposition of ``alive" and ``dead" corresponds to a region of state space that is dynamically inaccessible under measurement-like environmental interactions, because the macroscopic cat state is dynamically confined to a classical equivalence class by continual environmental monitoring. The paradox arises only if one assumes that such macroscopic superpositions are physically realizable, which they are not within this dynamics.

\subsection{Wigner's friend}

\emph{Question:}
How can one observer (the friend) record a definite measurement outcome, for example, the position of a particle, while another observer (Wigner) consistently describes the same system as being in a superposition?

\medskip

\emph{Answer:}
In this framework, the measuring device, the friend, and Wigner are all macroscopic systems interacting with the environment and evolving under {\bf (RM)}.
The joint state of the particle-device system (in one spatial dimension) has the product form $\varphi \otimes \Psi$, where $\varphi \in L_2(\mathbb{R})$ is the particle's state and $\Psi \in \widetilde M^{\sigma_d}_{1,1}$ is the state of the device.
Under measurement, the stochastic unitary dynamics drives the state into a definite equivalence class
$\{g_{c,\sigma_p}\} \otimes \{g_{d,\sigma_d}\} \in \widetilde M^{\sigma_p}_{1} \otimes \widetilde M^{\sigma_d}_{1}$,
corresponding to a definite measurement outcome. The device state $g_{d,\sigma_d}$ records this outcome through the mechanism described in the {\it Quantum-Classical Transition} section.

When a friend performs the measurement, the total system consists of the particle, the device, and the friend, with joint state $\varphi \otimes \Psi \otimes \Pi$, where $\Pi \in \widetilde M^{\sigma_f}_{1}$ represents the friend's macroscopic state.
The measurement interaction between the particle, the device, and the friend again drives the joint state into a single equivalence class
$\{g_{c,\sigma_p}\} \otimes \{g_{d,\sigma_d}\} \otimes \{g_{f,\sigma_f}\}
\in \widetilde M^{\sigma_p}_{1} \otimes \widetilde M^{\sigma_d}_{1} \otimes \widetilde M^{\sigma_f}_{1}$,
which is the only physically realized outcome.

When Wigner subsequently interacts with the composite system, the same stochastic unitary dynamics applies to the state
$\varphi \otimes \Psi \otimes \Pi \otimes \Phi$, where $\Phi \in \widetilde M^{\sigma_W}_{1}$ represents Wigner's macroscopic state.
The joint state after this interaction belongs to the equivalence class
$\{g_{c,\sigma_p}\} \otimes \{g_{d,\sigma_d}\} \otimes \{g_{f,\sigma_f}\} \otimes \{g_{W,\sigma_W}\}
\in \widetilde M^{\sigma_p}_{1} \otimes \widetilde M^{\sigma_d}_{1} \otimes \widetilde M^{\sigma_f}_{1} \otimes \widetilde M^{\sigma_W}_{1}$,
and Wigner necessarily records the same outcome $c$ for the particle's position as the friend.
No observer-dependent collapse, branching, or change of physical description occurs.
State reduction is an objective dynamical process determined by the evolution in state space, rather than by the perspective of a macroscopic observer.

This framework differs from Everettian and relational approaches in how outcome definiteness is obtained.
In Everettian accounts, linear Schr\"odinger evolution is taken to imply branching into multiple coexisting outcomes.
Here, by contrast, the stochastic unitary dynamics generated by random Hamiltonians leads to a single realized equivalence class in the classical submanifold, without invoking branching or parallel outcomes.

Similarly, unlike relational interpretations, outcome definiteness is not observer-relative.
All macroscopic systems, including measuring devices and observers, evolve under the same stochastic Schr\"odinger dynamics, and once the joint state reaches a classical equivalence class, all participants record the same outcome.
In this sense, state reduction is an objective dynamical process in state space rather than a perspectival or observer-dependent phenomenon.

\subsection{Decoherence}

\emph{Question:} What role does decoherence play in this framework, and how does it relate to state reduction and classical behavior?

\medskip

\emph{Answer:}
In standard accounts, decoherence explains the suppression of interference between macroscopically distinct states through entanglement with environmental degrees of freedom. While this accounts for the practical disappearance of off-diagonal terms in reduced density matrices, it does not by itself select a unique outcome, leaving the measurement problem unresolved.

In the present framework, decoherence emerges as a consequence of stochastic unitary evolution driven by random Hamiltonians in {\bf (RM)}, together with the identification of equivalence classes of detector-indistinguishable states. Environmental interactions rapidly drive the state toward the classical submanifold of state space, but it is the projection onto an equivalence class, rather than tracing over environmental degrees of freedom, that suppresses interference. Coherence is lost not merely because phases become inaccessible, but because states belonging to different equivalence classes are dynamically separated in the Fubini-Study geometry, with interference between them exponentially suppressed.

Environment-induced decoherence yields an effective mixture over classical alternatives, whereas the present approach yields an objective approach of the state to a specific equivalence class, corresponding to a definite measurement outcome. Decoherence therefore appears here not as an autonomous mechanism, but as a geometric and dynamical feature of state-space trajectories under {\bf (RM)}.
In this sense, decoherence, state reduction, and the quantum-classical transition are unified within a single stochastic unitary dynamics. Equivalence classes determine what counts as a classical record, while random Hamiltonians determine how and when such records emerge.

Note that when the initial state of a macroscopic body belongs to an equivalence class representing a point of the classical space or phase-space submanifold, subsequent evolution under {\bf (RM)}, together with continual environmental monitoring, assigns a vanishingly small probability to the state evolving into a superposition of states from different equivalence classes. In this sense, decoherence for macroscopic bodies becomes effectively trivial: the state remains confined to a single classical trajectory rather than undergoing coherent branching.

\subsection{Double-Slit Experiment}

\emph{Question:}  
How can a particle in the double-slit experiment display interference when its position is not measured at the slits, yet appear as a localized object upon measurement, thereby destroying interference? In such a process, does the particle pass through one slit or both slits?
\medskip

\emph{Answer:}  
At emission, the particle's state is well localized and lies on the classical phase-space submanifold $\widetilde M^{\sigma}_{3,3}$. In this regime, the particle behaves classically: its position is well defined, and its Schr\"odinger evolution reproduces Newtonian motion.

Upon interaction with the slit screen, with both slits open, the particle's state becomes a superposition of states localized near the individual slits. This means that the state moves away from the submanifold $\widetilde M^{\sigma}_{3}$, so that the distance (\ref{dist}) between the state and the submanifold increases. In this sense, the particle's path does not pass through either slit, since that would require the state to lie on $\widetilde M^{\sigma}_3$. Rather, upon interaction with the screen, the path of the state leaves the classical space submanifold and evolves through the full projective state space.

If the particle's position is measured at the slits, the stochastic dynamics described by {\bf (RM)} drives the state back to $\widetilde M^{\sigma}_3$. The particle is then detected at one of the slits, with probabilities determined by the Born rule. Once localized, the particle again follows a classical trajectory, in which case no interference pattern appears at the detection screen.

If no position measurement is performed at the slits, the state continues to evolve at a distance from $\widetilde M^{\sigma}_3$ as a superposition. When the particle reaches the detection screen, interaction with the screen induces a stochastic evolution under {\bf (RM)} that returns the state to $\widetilde M^{\sigma}_3$. In this case, the Born rule applied to the returning state yields the familiar interference pattern.

In this framework, the apparent wave-particle duality reflects motion in state space rather than the propagation of a physical wave in classical space. What appears as wave spreading corresponds to the state, that is, a point in state space, moving away from the classical submanifold, while localization corresponds to its stochastic return to $\widetilde M^{\sigma}_3$. Whenever the state is close to $\widetilde M^{\sigma}_3$ in the Fubini-Study-based metric (\ref{dist}), it behaves like a particle. Whenever the state moves away from $\widetilde M^{\sigma}_3$, it exhibits wave properties. 

Collapse is therefore not a sudden spatial contraction, but a dynamical transition in state space, realized as a motion of a point-state, governed by {\bf (RM)}, with outcome probabilities fixed by geometry.

\subsection{Nonlocality and EPR Correlations}

\emph{Question:}
How can quantum mechanics exhibit strong correlations between spatially separated systems, as in EPR and Bell-type experiments, without violating relativistic causality? Does the present framework require nonlocal dynamics?

\medskip

\emph{Answer:}
In the present framework, EPR correlations arise from the geometry of state space and the stochastic unitary dynamics generated by {\bf (RM)}, rather than from any nonlocal physical influence between particles in classical space.

An entangled two-particle system is represented by a single state evolving as a point in the joint projective state space $\mathbb{CP}^{L_2}$, based on the Hilbert space $L_2(\R) \otimes L_2(\R) \cong L_2(\mathbb{R}\times\mathbb{R})$ (here written for one spatial dimension). The random walk generated by {\bf (RM)} acts on this joint state as a whole. Transition probabilities depend only on Fubini-Study distances between states and satisfy the Born rule, thereby encoding correlations already present in the entangled state. 

There is no conflict with Bell's theorem, since the latter constrains local hidden-variable models formulated in spacetime, whereas the relevant variables here are global geometric properties of the joint quantum state evolving in state space. Accordingly, there is no motion in classical space and no signal or causal influence propagates between spatially separated particles during measurement.

When a position measurement is performed on one particle in a position-entangled state, so that the position of one particle determines the position of the other, the point representing the joint state dynamically approaches a particular equivalence class
$\{g_{c,\sigma_1}\otimes g_{d,\sigma_2}\}$ in the classical submanifold
$\R^2=\widetilde M^{\sigma_1}_{1}\otimes \widetilde M^{\sigma_2}_{1}$,
corresponding to a definite pair of classical position outcomes $c$ and $d$ for the two particles. Each outcome takes values in the one-dimensional classical space $\mathbb{R}$.
The entanglement present in the initial state, together with the validity of the Born rule for the random walk generated by {\bf (RM)}, implies that a state of this form is the only possible outcome of the measurement. In particular, if during the measurement process the state of the first particle becomes well defined and belongs to the class $\{g_{c,\sigma_1}\}$, then the state of the second particle must simultaneously become well defined and belong to the corresponding class $\{g_{d,\sigma_2}\}$.

The correlated outcome for the distant particle is not produced by a nonlocal action, but follows from the motion of the joint state toward a single equivalence class $\{g_{c,\sigma_1}\otimes g_{d,\sigma_2}\}$ determined by the initial entangled state and the measurement dynamics. The resulting Born-rule correlations thus emerge from the initial state of the pair (i.e., the initial point of the random walk), the geometry of the joint state space, and the homogeneity and isotropy of the stochastic dynamics, rather than from any superluminal influence.

The framework preserves parameter independence: measurement settings on one side do not affect the marginal outcome statistics on the other. At the same time, it allows for outcome dependence, exactly as required by quantum mechanics.
In this sense, the apparent nonlocality of EPR correlations reflects the inseparability of entangled states in state space, rather than any nonlocal dynamics in classical space. For a position-entangled state, this inseparability signifies that the initial entangled state lies outside the classical configuration space
$\R^2=\widetilde M^{\sigma_1}_{1}\otimes \widetilde M^{\sigma_2}_{1}$. 

The subsequent evolution toward
$\widetilde M^{\sigma_1}_{1}\otimes \widetilde M^{\sigma_2}_{1}$
remains unitary and local in time, i.e., Markovian, with the evolution at each instant generated by an independently drawn Hamiltonian. Moreover, the evolution is local in state space: infinitesimal time increments produce ininitesimal displacements in the Fubini-Study metric. 
Relativistic causality is preserved: measurement outcomes are correlated but cannot be used for superluminal signaling, since no motion of any kind occurs in classical space.

This perspective differs both from hidden-variable approaches, which posit additional nonlocal variables, and from interpretations that treat collapse as observer-dependent. In the present framework, correlations are objective properties of state-space trajectories generated by {\bf (RM)}. Measurement leads to the selection of a definite equivalence class of the joint system, yielding correlated outcomes without invoking nonlocal collapse or branching.
Accordingly, EPR correlations and Bell-inequality violations are fully compatible with a stochastic yet unitary measurement dynamics: quantum nonlocality is reinterpreted as a geometric feature of state space, its classical submanifolds, and the stochastic dynamics generated by {\bf (RM)}.

\section{Conceptual Implications and Further Paradoxes}

The framework developed above provides a unified dynamical account of measurement, state reduction, and the quantum-classical transition, based on stochastic yet unitary evolution in state space. In this section, we briefly address several foundational issues and paradoxes that are often treated as independent problems, and show how they are naturally reformulated and addressed within the same underlying mechanism.

\subsection{What Is Real in a Superposition}

In the present framework, a quantum state is represented by a point in the projective state space $\mathbb{CP}^{L_2}$, and physical evolution corresponds to a path of this point under Schr\"odinger dynamics generated by a random Hamiltonian, as specified in {\bf (RM)}. A superposition of states from the equivalence classes under consideration does not represent the coexistence of multiple classical realities; rather, it corresponds to a state lying away from the classical configuration-space or phase-space submanifolds of $\mathbb{CP}^{L_2}$.
Classical properties become well defined only when the state approaches these submanifolds. In this sense, what is physically real at any given time is the path traced by the state in state space itself. The apparent ambiguity of superpositions arises only when one attempts to ascribe classical attributes to states that do not lie on classical submanifolds.

This perspective removes the need to interpret superpositions as describing multiple simultaneously realized outcomes, while also avoiding hidden variables. Superpositions in this framework are not merely statistical bookkeeping or information-bearing mathematical constructs. They represent genuine physical quantum states of systems, as already evidenced by their ability to encode physical quantities such as position and momentum in special cases. However, they do not correspond to definite classical configurations until the dynamics drives them into the appropriate equivalence classes. 

Adopting this ontological reading is not required for the formalism to operate, but it provides a coherent and economical picture in which measurement, classicality, and state reduction all arise from a single dynamical mechanism and the geometry of state space.

\subsection{Preferred Basis and Classical Observables}

A longstanding difficulty in quantum foundations is the preferred-basis problem: why measurement outcomes appear in particular bases (most notably position) rather than in arbitrary superpositions. In standard environment-induced decoherence, this issue is addressed by invoking environment-selected pointer states, typically in a model-dependent way.

In the present framework, the step distribution of the random walk in {\bf (RM)} is homogeneous and isotropic in state space. The dynamics generated by {\bf (RM)} can drive an initial state toward classical configuration space, classical phase space, or other physically meaningful (i.e., observable-related) submanifolds of state space, with probabilities determined solely by the Fubini-Study distance to the corresponding end states. In this sense, the dynamics itself does not single out a preferred basis.

Position measurement nevertheless plays a distinguished role, since realistic measuring devices ultimately register position, often only indirectly. Other observables are accessed by correlating them with position through the design of the measurement apparatus. For example, when measuring the momentum of a charged particle, a magnetic spectrometer maps momentum values to spatial locations on a detection screen. Measuring position therefore imposes equivalence relations determined by spatial resolution, thereby defining the appropriate classical submanifolds, in this case, the manifold of approximate eigenstates of the momentum operator. This latter manifold may be constructed mathematically by applying the Fourier transform to $M^{\sigma}_3$.

No additional assumptions are required to select a basis. Measurement, together with its possible outcomes, is determined by the geometry of state space and its classical submanifolds, the properties of the random walk in {\bf (RM)}, the design of the measuring device or the properties of the environment, and the physical constraints of measurement. An observable is classical precisely when the state lies on the corresponding classical submanifold.

\subsection{Quantum Zeno Effect}

The quantum Zeno effect is usually described as the inhibition of evolution caused by frequent measurements, resulting in repeated application of the projection postulate. In the present framework, the Born rule is derived dynamically from the random walk in {\bf (RM)}, which means that projection is incorporated into the dynamics of the state itself. It follows that the Zeno effect arises from the dynamics alone, with no need for an additional projection postulate.

In particular, consider a position measurement of a particle. As the time between stochastic kicks of the random walk in {\bf (RM)} decreases, the probability that the state revisits an equivalence class within a fixed, arbitrarily small time interval approaches unity. Frequent environmental or device-induced measurement generates the corresponding stochastic evolution, while recording confirms that the state satisfies the equivalence-class condition. Each such confirmation effectively restarts the stochastic evolution from the corresponding location within the equivalence class. The process that leads the state to these locations plays the role traditionally attributed to state projection.  As a result, the probability that the state escapes a small neighborhood of the equivalence class becomes vanishingly small.

The resulting suppression of transitions therefore follows from the statistical properties of the stochastic dynamics, the geometry of state space and its classical submanifolds, together with the recording of the state's locations on the classical space submanifold. As discussed earlier, when a free Hamiltonian is added to the Hamiltonian in {\bf (RM)}, the state is confined, under appropriate conditions, to a small neighborhood of the classical space submanifold and undergoes Newtonian motion there.

Thus, the quantum Zeno effect is reinterpreted as a dynamical stabilization of state-space paths, produced by repeated measurement-induced unitary evolution under {\bf (RM)}, which suppresses transitions of the state. Recording the state's membership in an equivalence class plays no dynamical role, but merely confirms that the particle's position is well defined and either remains fixed or evolves in accordance with Newtonian dynamics.

\subsection{Irreversibility and the Arrow of Time}

The stochastic unitary dynamics introduced in {\bf (RM)} gives rise to an arrow of time. This arrow does not originate from a single mechanism, but from the combined effect of three distinct and logically complementary ingredients: (i) the properties of stochastic evolution on high-dimensional state space,
(ii) the properties of random Hamiltonians drawn from the Gaussian Unitary Ensemble, and
(iii) the use of equivalence classes and recording in the description of measurement outcomes.

First, stochastic Schr\"odinger evolution driven by random Hamiltonians produces irreversible behavior at the level of typical state-space trajectories. In infinite-dimensional (or sufficiently high-dimensional) projective Hilbert space, random unitary evolution almost surely carries a state away from any given neighborhood of its initial position. Exact or approximate recurrences have vanishing probability. This establishes a minimal, observer-independent arrow of time: forward evolution explores an ever-increasing region of state space, while backward reconstruction of a trajectory is overwhelmingly unlikely. This irreversibility in state space is a purely dynamical effect, independent of measurement, coarse-graining, or recording.

Second, the choice of ensemble in {\bf (RM)} plays a crucial role. On the one hand, the Gaussian Unitary Ensemble is required to ensure the universal validity of the Born rule for transitions between arbitrary quantum states under {\bf (RM)}. On the other hand, Hamiltonians drawn from the GUE explicitly break time-reversal symmetry through their complex matrix elements, leading to irreversible scrambling of relative phases in the state. Since time reversal corresponds to complex conjugation, the state does not retrace its original path in state space. It follows that the GUE induces a strong form of microscopic irreversibility by destroying phase correlations and enforcing isotropy of the stochastic dynamics in complex state space.

Third, irreversibility is amplified by the introduction of equivalence classes of detector-indistinguishable states. Each equivalence class contains infinitely many mutually orthogonal states and retains only experimentally accessible information, such as expectation values within finite resolution. Identifying a physical state with an equivalence class therefore constitutes an intrinsic coarse-graining in state space. Information about relative phases and microscopic degrees of freedom within a class is discarded by construction and cannot be recovered. As a result, evolution described in terms of equivalence classes is fundamentally irreversible, even though the underlying dynamics remains unitary.

Finally, recording plays an important but purely informational role. Whenever the evolving state enters an equivalence class of the classical submanifold, this fact may be recorded by a measuring device or the environment. Recording does not affect the dynamics of the state; rather, each recorded outcome becomes the center of a new probability distribution for subsequent evolution. Repeated recording therefore suppresses large excursions away from the classical submanifold and stabilizes classical trajectories. The past is fixed by records, while the future remains probabilistic. This asymmetry between past and future completes the emergence of a physical arrow of time.

\section{Summary}

We have presented a unified dynamical framework in which stochastic but unitary evolution in state space accounts for measurement, state reduction, and the quantum-classical transition. The approach is based on three key elements: the geometry of projective state space equipped with the Fubini-Study metric and its classical submanifolds; stochastic unitary dynamics generated by time-dependent random Hamiltonians as postulated in {\bf (RM)}; and equivalence classes of detector-indistinguishable states encoding the finite resolution of realistic measurements.

Within this framework, measurement in both quantum and classical regimes is described by the same stochastic unitary dynamics, operating in different parameter ranges. Classical configuration space and phase space emerge as distinguished submanifolds of state space, and classical behavior corresponds to dynamical confinement of states to neighborhoods of these submanifolds. State reduction arises as a consequence of this stochastic process, corresponding to the dynamical approach of the state toward the classical configuration-space submanifold. In this setting, the Born rule governs transition probabilities for microscopic systems, while normal probability distributions emerge for the measured positions of macroscopic objects.

The framework admits experimental tests through which its internal consistency and physically relevant parameter ranges may be assessed \cite{KryukovPhysicsA}. In particular, deviations from idealized measurement dynamics, the rate of approach to classical submanifolds, and the dependence of effective collapse behavior on system size and environmental coupling provide potential avenues for comparison with interferometric, continuous-measurement, and mesoscopic experiments. These and related experimental tests will provide a conclusive assessment of the validity of conjecture {\bf (RM)} and of the dynamical consequences explored in this work.

\end{document}